\documentclass{article}

     \usepackage[preprint]{neurips_2020}

\usepackage[utf8]{inputenc} % allow utf-8 input
\usepackage[T1]{fontenc}    % use 8-bit T1 fonts
\usepackage{hyperref}       % hyperlinks
\usepackage{url}            % simple URL typesetting
\usepackage{booktabs}       % professional-quality tables
\usepackage{amsfonts}       % blackboard math symbols
\usepackage{nicefrac}       % compact symbols for 1/2, etc.
\usepackage{microtype}      % microtypography
\usepackage{subfig}

\usepackage{natbib}
\usepackage{graphicx}

\usepackage{amsfonts}
\usepackage{amsmath}
\usepackage{todo}
\usepackage{mathtools} 
\usepackage{extarrows} 
\usepackage{color}
\usepackage{wrapfig}
\usepackage{makecell}
\usepackage{url}

\usepackage{color}
\usepackage{amssymb}

\usepackage{pgfplots}
\usepackage{filecontents}
\makeatletter
\@namedef{ver@everyshi.sty}{}
\makeatother
\usepackage{tikz}
\usepackage{epstopdf}
\usepackage{subfig}
\usepackage{makecell}
\usepackage{booktabs}
\usepackage[bottom]{footmisc}

\definecolor{myred}{RGB}{200,100,0}
\definecolor{myblue}{RGB}{10,10,200}
\definecolor{mygreen}{RGB}{10,150,100}
\definecolor{myorange}{RGB}{255,165,0}

\DeclareMathOperator*{\argmin}{arg\,min}

\makeatletter
\renewcommand{\paragraph}{%
	\@startsection{paragraph}{4}%
	{\z@}{1.25ex \@plus 1ex \@minus .2ex}{-1em}%
	{\normalfont\normalsize\bfseries}%
}
\makeatother

\gdef\useCroppedImages{1}

\gdef\cropInsets{0}
\usepackage[export]{adjustbox}
\usepackage{calc}
\usepackage{tabularx}
\usepackage{tikz}
\usepackage{xstring}
\usepackage{float}

\newlength\beautyHeight
\newlength\beautyPixWidth
\newlength\beautyPixHeight
\newlength\insetvsep
\gdef\useInsetA{0}
\gdef\useInsetB{0}
\gdef\useInsetC{0}

\newcommand{\setInset}[6]{%
	\expandafter\gdef\csname useInset#1\endcsname{1}%
	\expandafter\gdef\csname inset#1Color\endcsname{#2}%
	\expandafter\gdef\csname crop#1X\endcsname{#3}%
	\expandafter\gdef\csname crop#1Y\endcsname{#4}%
	\expandafter\gdef\csname crop#1W\endcsname{#5}%
	\expandafter\gdef\csname crop#1H\endcsname{#6}%
}

\newcommand{\unsetInset}[1]{%
	\expandafter\gdef\csname useInset#1\endcsname{0}%
}

\newcommand{\addBeautyCrop}[8]{%
	\pdfpxdimen=\dimexpr 1 in/72\relax
	\def\beauty{%
		\let\cropR\relax%
		\let\cropB\relax%
		\newlength\cropR%
		\newlength\cropB%
		\setlength\cropR{{#3 px}-{#5 px}-{#7 px}}%
		\setlength\cropB{{#4 px}-{#6 px}-{#8 px}}%
		\sbox0{\includegraphics[width=#2\textwidth,trim={#5px {\cropB} {\cropR} #6px},clip]{#1}}%
		\begin{tikzpicture}
		\node[anchor=north west,inner sep=0] at (0,0) {\usebox0};
		\begin{scope}[x=\wd0/#7, y=\ht0/#8]
		\if\useInsetA1{
			\draw[\insetAColor,thick] (\cropAX-#5,-\cropAY+#6) rectangle + (\cropAW,-\cropAH);
		}\fi
		\if\useInsetB1{
			\draw[\insetBColor,thick] (\cropBX-#5,-\cropBY+#6) rectangle + (\cropBW,-\cropBH);
		}\fi
		\if\useInsetC1{
			\draw[\insetCColor,thick] (\cropCX-#5,-\cropCY+#6) rectangle + (\cropCW,-\cropCH);
		}\fi
		\end{scope}
		\end{tikzpicture}
	}%
	\setlength\beautyHeight{\heightof{\beauty}}%
	\setlength\beautyPixWidth{#3px}%
	\setlength\beautyPixHeight{#4px}%
	\global\beautyHeight=\beautyHeight%
	\global\beautyPixWidth=\beautyPixWidth%
	\global\beautyPixHeight=\beautyPixHeight%
	\begin{adjustbox}{valign=t}
		\beauty
	\end{adjustbox}
}

\newcommand{\addBeauty}[4]{%
	\addBeautyCrop{#1}{#2}{#3}{#4}{0}{0}{#3}{#4}%
}

\newcommand{\trimInset}[6]{%
	\let\cropR\relax%
	\let\cropB\relax%
	\newlength\cropR%
	\newlength\cropB%
	\setlength\cropR{{\beautyPixWidth}-{#3 px}-{#5 px}}%
	\setlength\cropB{{\beautyPixHeight}-{#4 px}-{#6 px}}%
	\color{#2}%
	\fbox{\includegraphics[width=1\linewidth,trim={{#3 px} {\cropB} {\cropR} {#4 px}},clip]{#1}}%
}

\newcommand{\addInset}[2]{%
	\color{#2}%
	\fbox{\includegraphics[width=1\linewidth]{#1}}%
}

\newcommand{\auxtimes}{x}
\newcommand{\auxplus}{+}
\newcommand{\auxspace}{ }

\newcommand{\addInsets}[2][1]{%
	\begin{adjustbox}{valign=t}
		\StrSubstitute{#2}{.}{-}[\baseFileName]
		\begin{adjustbox}{totalheight=#1\beautyHeight,tabular={c}}
			\if\useInsetA1%
			\def\cropfile{\baseFileName-\cropAW\auxtimes\cropAH\auxplus\cropAX\auxplus\cropAY}
			\if\cropInsets1
			\immediate\write18{convert #2 +repage -crop \cropAW\auxtimes\cropAH\auxplus\cropAX\auxplus\cropAY\auxspace \cropfile.png}
			\fi
			\if\useCroppedImages1
			\addInset{\cropfile.png}{\insetAColor}
			\else
			\trimInset{#2}{\insetAColor}{\cropAX}{\cropAY}{\cropAW}{\cropAH}%
			\fi%
			\fi%
			\if\useInsetB1%
			\if\useInsetA1\\[\insetvsep]\fi%
			\def\cropfile{\baseFileName-\cropBW\auxtimes\cropBH\auxplus\cropBX\auxplus\cropBY}
			\if\cropInsets1
			\immediate\write18{convert #2 -crop \cropBW\auxtimes\cropBH\auxplus\cropBX\auxplus\cropBY\auxspace \cropfile.png}
			\fi
			\if\useCroppedImages1
			\addInset{\cropfile.png}{\insetBColor}
			\else
			\trimInset{#2}{\insetBColor}{\cropBX}{\cropBY}{\cropBW}{\cropBH}%
			\fi%
			\fi%
			\if\useInsetC1%
			\if\useInsetB1\\[\insetvsep]\fi%
			\def\cropfile{\baseFileName-\cropCW\auxtimes\cropCH\auxplus\cropCX\auxplus\cropCY}
			\if\cropInsets1
			\immediate\write18{convert #2 -crop \cropCW\auxtimes\cropCH\auxplus\cropCX\auxplus\cropCY\auxspace \cropfile.png}
			\fi
			\if\useCroppedImages1
			\addInset{\cropfile.png}{\insetCColor}
			\else
			\trimInset{#2}{\insetCColor}{\cropCX}{\cropCY}{\cropCW}{\cropCH}%
			\fi%
			\fi%
		\end{adjustbox}
	\end{adjustbox}
}

\newcommand{\addHorInsets}[2][1]{%
	\begin{adjustbox}{valign=t}
		\StrSubstitute{#2}{.}{-}[\baseFileName]
		\begin{adjustbox}{totalheight=#1\beautyHeight,tabular={c}}
			\if\useInsetA1%
			\def\cropfile{\baseFileName-\cropAW\auxtimes\cropAH\auxplus\cropAX\auxplus\cropAY}
			\if\cropInsets1
			\immediate\write18{convert #2 -crop \cropAW\auxtimes\cropAH\auxplus\cropAX\auxplus\cropAY\auxspace \cropfile.png}
			\fi
			\if\useCroppedImages1
			\addInset{\cropfile.png}{\insetAColor}
			\else
			\trimInset{#2}{\insetAColor}{\cropAX}{\cropAY}{\cropAW}{\cropAH}%
			\fi%
			\fi%
			\if\useInsetB1%
			\def\cropfile{\baseFileName-\cropBW\auxtimes\cropBH\auxplus\cropBX\auxplus\cropBY}
			\if\cropInsets1
			\immediate\write18{convert #2 -crop \cropBW\auxtimes\cropBH\auxplus\cropBX\auxplus\cropBY\auxspace \cropfile.png}
			\fi
			\if\useCroppedImages1
			\addInset{\cropfile.png}{\insetBColor}
			\else
			\trimInset{#2}{\insetBColor}{\cropBX}{\cropBY}{\cropBW}{\cropBH}%
			\fi%
			\fi%
			\if\useInsetC1%
			\def\cropfile{\baseFileName-\cropCW\auxtimes\cropCH\auxplus\cropCX\auxplus\cropCY}
			\if\cropInsets1
			\immediate\write18{convert #2 -crop \cropCW\auxtimes\cropCH\auxplus\cropCX\auxplus\cropCY\auxspace \cropfile.png}
			\fi
			\if\useCroppedImages1
			\addInset{\cropfile.png}{\insetCColor}
			\else
			\trimInset{#2}{\insetCColor}{\cropCX}{\cropCY}{\cropCW}{\cropCH}%
			\fi%
			\fi%
		\end{adjustbox}
	\end{adjustbox}
}

\pgfkeys{/tikz/.cd,
	zoombox paths/.style={
		draw=orange,
		very thick
	},
	black and white/.is choice,
	black and white/.default=static,
	black and white/static/.style={ 
		draw=white,   
		zoombox paths/.append style={
			draw=white,
			postaction={
				draw=black,
				loosely dashed
			}
		}
	},
	black and white/static/.code={
		\gdef\patternnumber{1}
	},
	black and white/cycle/.code={
		\blackandwhitecycletrue
		\gdef\patternnumber{1}
	},
	black and white pattern/.is choice,
	black and white pattern/0/.style={},
	black and white pattern/1/.style={    
		draw=white,
		postaction={
			draw=black,
			dash pattern=on 2pt off 2pt
		}
	},
	black and white pattern/2/.style={    
		draw=white,
		postaction={
			draw=black,
			dash pattern=on 4pt off 4pt
		}
	},
	black and white pattern/3/.style={    
		draw=white,
		postaction={
			draw=black,
			dash pattern=on 4pt off 4pt on 1pt off 4pt
		}
	},
	black and white pattern/4/.style={    
		draw=white,
		postaction={
			draw=black,
			dash pattern=on 4pt off 2pt on 2 pt off 2pt on 2 pt off 2pt
		}
	},
	zoomboxarray inner gap/.initial=5pt,
	zoomboxarray columns/.initial=2,
	zoomboxarray rows/.initial=2,
	subfigurename/.initial={},
	figurename/.initial={zoombox},
	zoomboxarray/.style={
		execute at begin picture={
			\begin{scope}[
				spy using outlines={%
					zoombox paths,
					width=\imagewidth / \pgfkeysvalueof{/tikz/zoomboxarray columns} - (\pgfkeysvalueof{/tikz/zoomboxarray columns} - 1) / \pgfkeysvalueof{/tikz/zoomboxarray columns} * \pgfkeysvalueof{/tikz/zoomboxarray inner gap} -\pgflinewidth,
					height=\imageheight / \pgfkeysvalueof{/tikz/zoomboxarray rows} - (\pgfkeysvalueof{/tikz/zoomboxarray rows} - 1) / \pgfkeysvalueof{/tikz/zoomboxarray rows} * \pgfkeysvalueof{/tikz/zoomboxarray inner gap}-\pgflinewidth,
					magnification=3,
					every spy on node/.style={
						zoombox paths
					},
					every spy in node/.style={
						zoombox paths
					}
				}
				]
			},
			execute at end picture={
			\end{scope}
			\node at (image.north) [anchor=north,inner sep=0pt] {\subcaptionbox{\label{\pgfkeysvalueof{/tikz/figurename}-image}}{\phantomimage}};
			\node at (zoomboxes container.north) [anchor=north,inner sep=0pt] {\subcaptionbox{\label{\pgfkeysvalueof{/tikz/figurename}-zoom}}{\phantomimage}};
			\gdef\patternnumber{0}
		},
		spymargin/.initial=0.5em,
		zoomboxes xshift/.initial=1,
		zoomboxes right/.code=\pgfkeys{/tikz/zoomboxes xshift=1},
		zoomboxes left/.code=\pgfkeys{/tikz/zoomboxes xshift=-1},
		zoomboxes yshift/.initial=0,
		zoomboxes above/.code={
			\pgfkeys{/tikz/zoomboxes yshift=1},
			\pgfkeys{/tikz/zoomboxes xshift=0}
		},
		zoomboxes below/.code={
			\pgfkeys{/tikz/zoomboxes yshift=-1},
			\pgfkeys{/tikz/zoomboxes xshift=0}
		},
		caption margin/.initial=4ex,
	},
	adjust caption spacing/.code={},
	image container/.style={
		inner sep=0pt,
		at=(image.north),
		anchor=north,
		adjust caption spacing
	},
	zoomboxes container/.style={
		inner sep=0pt,
		at=(image.north),
		anchor=north,
		name=zoomboxes container,
		xshift=\pgfkeysvalueof{/tikz/zoomboxes xshift}*(\imagewidth+\pgfkeysvalueof{/tikz/spymargin}),
		yshift=\pgfkeysvalueof{/tikz/zoomboxes yshift}*(\imageheight+\pgfkeysvalueof{/tikz/spymargin}+\pgfkeysvalueof{/tikz/caption margin}),
		adjust caption spacing
	},
	calculate dimensions/.code={
		\pgfpointdiff{\pgfpointanchor{image}{south west} }{\pgfpointanchor{image}{north east} }
		\pgfgetlastxy{\imagewidth}{\imageheight}
		\global\let\imagewidth=\imagewidth
		\global\let\imageheight=\imageheight
		\gdef\columncount{1}
		\gdef\rowcount{1}
		
	},
	image node/.style={
		inner sep=0pt,
		name=image,
		anchor=south west,
		append after command={
			[calculate dimensions]
			node [image container,subfigurename=\pgfkeysvalueof{/tikz/figurename}-image] {\phantomimage}
			node [zoomboxes container,subfigurename=\pgfkeysvalueof{/tikz/figurename}-zoom] {\phantomimage}
		}
	},
	color code/.style={
		zoombox paths/.append style={draw=#1}
	},
	connect zoomboxes/.style={
		spy connection path={\draw[draw=none,zoombox paths] (tikzspyonnode) -- (tikzspyinnode);}
	},
	help grid code/.code={
		\begin{scope}[
			x={(image.south east)},
			y={(image.north west)},
			font=\footnotesize,
			help lines,
			overlay
			]
			\foreach \x in {0,1,...,9} { 
				\draw(\x/10,0) -- (\x/10,1);
				\node [anchor=north] at (\x/10,0) {0.\x};
			}
			\foreach \y in {0,1,...,9} {
				\draw(0,\y/10) -- (1,\y/10);                        \node [anchor=east] at (0,\y/10) {0.\y};
			}
		\end{scope}    
	},
	help grid/.style={
		append after command={
			[help grid code]
		}
	},
}

\newcommand\phantomimage{%
	\phantom{%
		\rule{\imagewidth}{\imageheight}%
	}%
}
\newcommand\zoombox[2][]{
	\begin{scope}[zoombox paths]
		\pgfmathsetmacro\xpos{
			(\columncount-1)*(\imagewidth / \pgfkeysvalueof{/tikz/zoomboxarray columns} + \pgfkeysvalueof{/tikz/zoomboxarray inner gap} / \pgfkeysvalueof{/tikz/zoomboxarray columns} ) + \pgflinewidth
		}
		\pgfmathsetmacro\ypos{
			(\rowcount-1)*( \imageheight / \pgfkeysvalueof{/tikz/zoomboxarray rows} + \pgfkeysvalueof{/tikz/zoomboxarray inner gap} / \pgfkeysvalueof{/tikz/zoomboxarray rows} ) + 0.5*\pgflinewidth
		}
		\edef\dospy{\noexpand\spy [
			#1,
			zoombox paths/.append style={
				black and white pattern=\patternnumber
			},
			every spy on node/.append style={#1},
			x=\imagewidth,
			y=\imageheight
			] on (#2) in node [anchor=north west] at ($(zoomboxes container.north west)+(\xpos pt,-\ypos pt)$);}
		\dospy
		\pgfmathtruncatemacro\pgfmathresult{ifthenelse(\columncount==\pgfkeysvalueof{/tikz/zoomboxarray columns},\rowcount+1,\rowcount)}
		\global\let\rowcount=\pgfmathresult
		\pgfmathtruncatemacro\pgfmathresult{ifthenelse(\columncount==\pgfkeysvalueof{/tikz/zoomboxarray columns},1,\columncount+1)}
		\global\let\columncount=\pgfmathresult
		\ifblackandwhitecycle
		\pgfmathtruncatemacro{\newpatternnumber}{\patternnumber+1}
		\global\edef\patternnumber{\newpatternnumber}
		\fi
	\end{scope}
}

\usepackage[skins]{tcolorbox} % for the teaser
\usepackage{shadowtext}
\shadowrgb{0,0,0}
\shadowoffset{0.5pt}

\title{Blind Image Restoration with Flow Based Priors}

\DeclareMathOperator*{\argmax}{arg\,max}
\newcommand{\printfnsymbol}[1]{%
	\textsuperscript{\@fnsymbol{#1}}%
}

\author{%
	 Leonhard Helminger$^{1,}$\footnotemark[1]\quad Michael Bernasconi$^{1,}$\thanks{equal contribution}\quad Abdelaziz Djelouah$^2$\\ \\
	\textbf{Markus Gross$^1$\quad Christopher Schroers$^2$}\\
	\\
	$^1$Department of Computer Science\quad \quad \quad \quad  $^2$DisneyResearch|Studios\\
	~~~~~ETH Zurich, Switzerland\quad \quad \quad \quad \quad \quad \quad Zurich, Switzerland\\
}

\begin{document}

\maketitle

\begin{abstract}
Image restoration has seen great progress in the last years thanks to the advances 
in deep neural networks. Most of these existing techniques are trained using 
full supervision with suitable image pairs to tackle a specific degradation. 
However, in a blind setting with unknown degradations this is not possible 
and a good prior remains crucial. Recently, neural network based approaches have 
been proposed to model such priors by leveraging either denoising autoencoders 
or the implicit regularization captured by the neural network structure itself.
In contrast to this, we propose using normalizing flows to model 
the distribution of the target content and to use this as a prior in a maximum 
a posteriori (MAP) formulation. By expressing the MAP optimization process in 
the latent space through the learned bijective mapping, we are able to obtain 
solutions through gradient descent. To the best of our knowledge, this is the first 
work that explores normalizing flows as prior in image enhancement problems. 
Furthermore, we present experimental results for a number of different degradations 
on data sets varying in complexity and show competitive results 
when comparing with the deep image prior approach.
\end{abstract}

\section{Introduction}
% 1) Which problem do we address and why is it important.
In today's digitized world, there is an increased demand 
to process existing older content. Examples are the archival of photo prints~\citep{Photoscan}
for more reliable long-term data storage, preparing
heritage footage~\citep{AmericaInColor} for more engaging documentaries, and making classic films and existing catalog contents available to large new audiences through streaming services.
This old content is however often in low quality and may be deteriorated 
in complex ways, which creates a need for \emph{blind} image restoration 
methods that are generic and able to address a wide 
range of possibly combined degradations.
Blind image restoration can be formulated as solving 
the following energy minimization problem:
%%%
%
\begin{equation}
\label{eq:problem}
\mathbf{x}^\star = 
\argmin_\mathbf{x} \left[ \mathcal{L}_{\text{data}}\left(\hat{\mathbf{x}},~\mathbf{x}\right) 
+  \mathcal{L}_{\text{reg}}(\mathbf{x}) \right]\;,
\end{equation}
%%%
%
where $\hat{\mathbf{x}}$ is the observed image and $\mathbf{x}^\star$ 
the restored image to be estimated. The first term, $\mathcal{L}_{\text{data}}$, 
is a data fidelity term which can be problem dependent and ensures 
that the solution agrees with the observation; 
the second term, $\mathcal{L}_{\text{reg}}(\mathbf{x})$,
is a regularizer that typically encodes certain smoothness assumptions on the expected solution and thus pushes it to lie within a given space. 
From a Bayesian viewpoint, the posterior distribution 
of the restored image is 
$p (\mathbf{x} | \hat{ \mathbf{x} })\propto p (\hat{ \mathbf{x} } | \mathbf{x}) p(\mathbf{x}) $.
This allows rewriting the above restoration problem into the following equivalent 
maximum a posteriori (MAP) estimate:
%%%
\begin{equation}
\label{eq:map_problem}
\mathbf{x}^\star =  \argmax_\mathbf{x} ~\log(p (\mathbf{x} | \hat{ \mathbf{x} }) )=
\argmax_\mathbf{x} ~\underbrace{log \left(p \left( \hat{ \mathbf{x} } | \mathbf{x} \right) \right)}_\text{data} 
+ \underbrace{\log p(\mathbf{x})}_\text{reg} \;,
\end{equation}
%%%

% 2) What other people do.
which makes it more explicit that the regularizer should model prior knowledge about the unknown solution.
Many handcrafted priors have been proposed reflecting 
desired properties based on total variation~\citep{TVrudin1992},
gradient sparsity~\citep{sparsityfergus2006} or the dark pixel prior~\citep{he2010single}.
More recently, learning based priors have been explored,
in particular the usage of denoising autoencoders (DAEs)
as regularizers for inverse imaging problems~\citep{meinhardt2017learning}.
Building on DAEs, \cite{bigdeli2017deep} propose to use 
a Gaussian smoothed natural image distribution as prior.
In a different direction, \cite{ulyanov2018deep} 
showed that an important part of the image statistics is 
captured by the structure of a convolutional image generator even
independent of any learning.

\begin{figure}[t]
	\hspace{-0.65cm}
	\small
	\unsetInset{A}
	\unsetInset{B}
	\unsetInset{C}
	\begin{tabular}{c@{\hspace{2pt}}c@{\hspace{2pt}}c@{\hspace{1.5pt}}c@{\hspace{1.5pt}}c}
		
		\textbf{Degraded} &  \textbf{Ours} & \textbf{Ours} &  \textbf{\footnotesize\cite{ulyanov2018deep}} & \textbf{GT}\\
		%%-----------------------------
		%% F16
		%%-----------------------------
		%\setInset{C}{blue}{200}{310}{75}{75}
		\makecell[cc]{\addBeauty{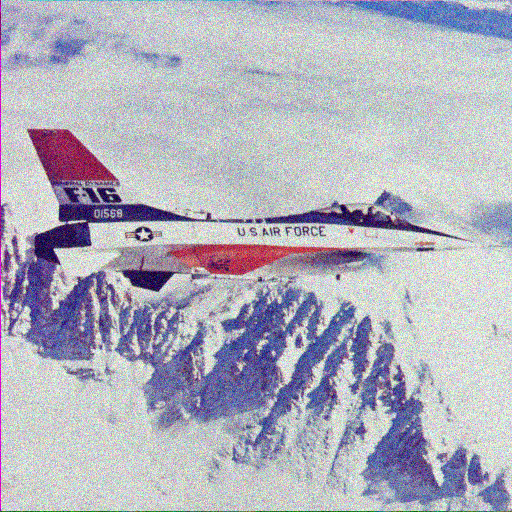}{0.205}{512}{512}}
		&
		\setInset{A}{red}{220}{220}{125}{65}
		\setInset{B}{cyan}{50}{175}{75}{50}
		\makecell[cc]{\addBeauty{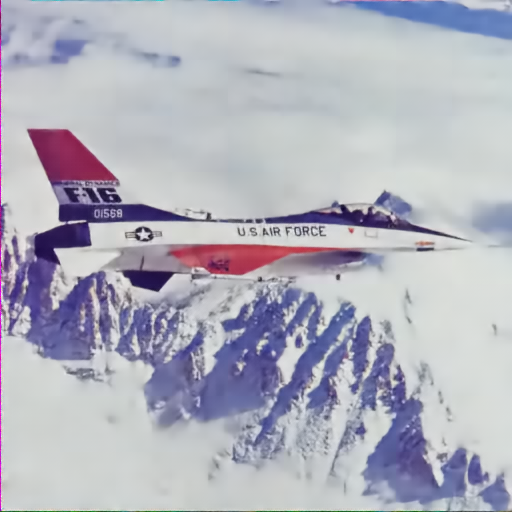}{0.205}{512}{512}}
		& 
		\makecell[cc]{\addInsets[1.01]{comparison_dip/ours/F16.png}}
		%& 
		%\makecell[cc]{\addInsets[1.01]{comparison_dip/degraded/F16.png}}
		& 
		\makecell[cc]{\addInsets[1.01]{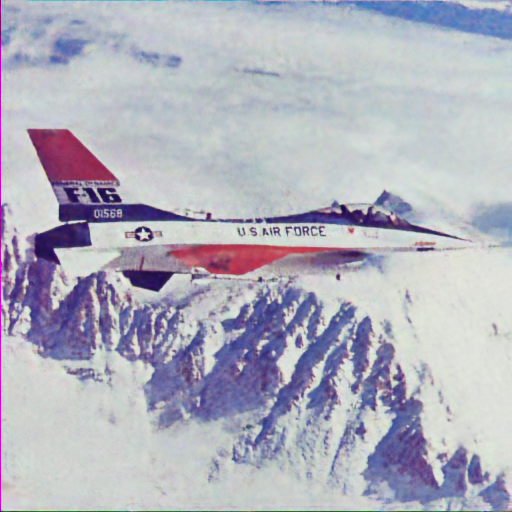}}	
		& 
		\makecell[cc]{\addInsets[1.01]{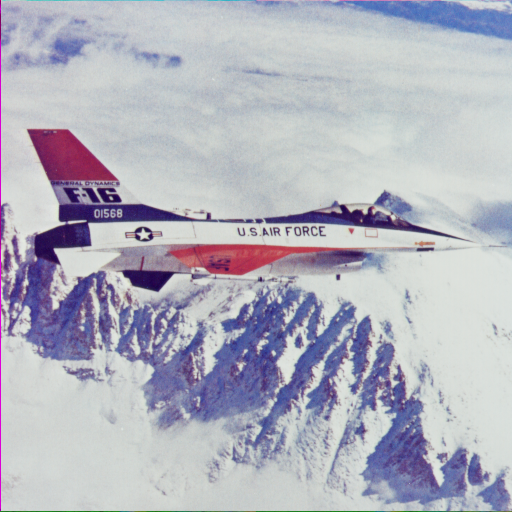}}		\\
		&&&&\\
		%%-----------------------------
		%% Kate
		%%-----------------------------
		\unsetInset{A}
		\unsetInset{B}
		\makecell[cc]{\addBeauty{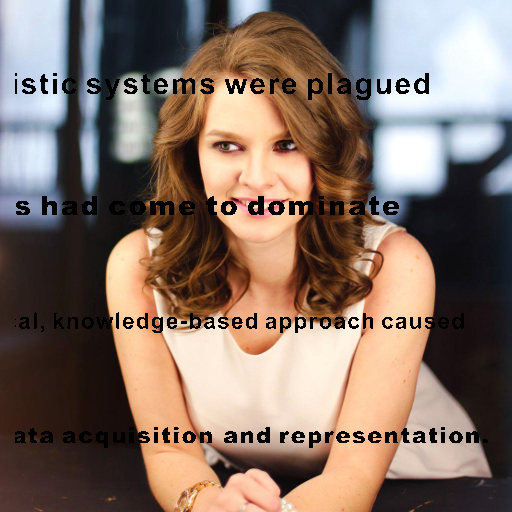}{0.2}{512}{512}}
		& 
		\setInset{A}{blue}{230}{190}{60}{30}
		\setInset{B}{green}{200}{200}{30}{20}
		\makecell[cc]{\addBeauty{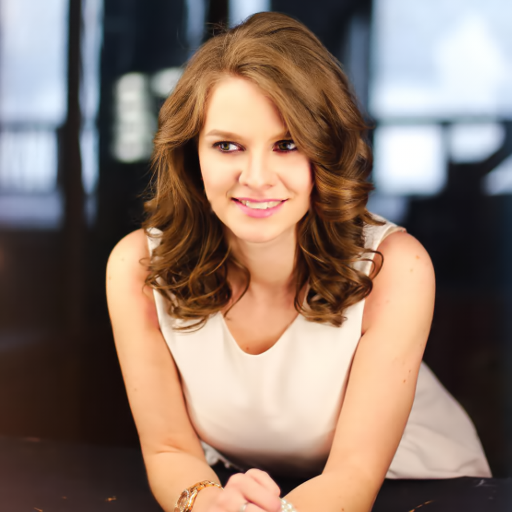}{0.2}{512}{512}}
		& 
		\makecell[cc]{\addInsets[1]{comparison_dip/ours/kate.png}}
		& 
		\makecell[cc]{\addInsets[1]{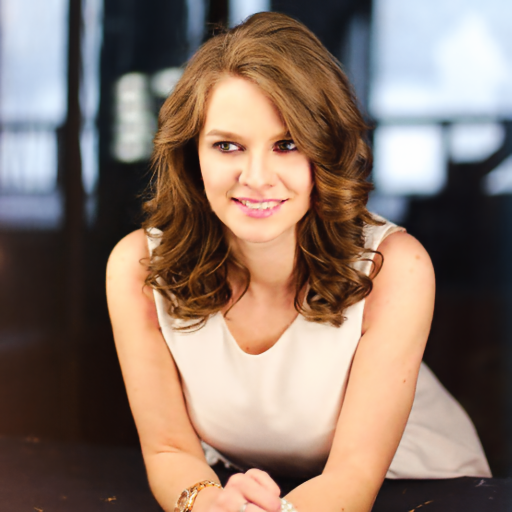}}	
		& 
		\makecell[cc]{\addInsets[1]{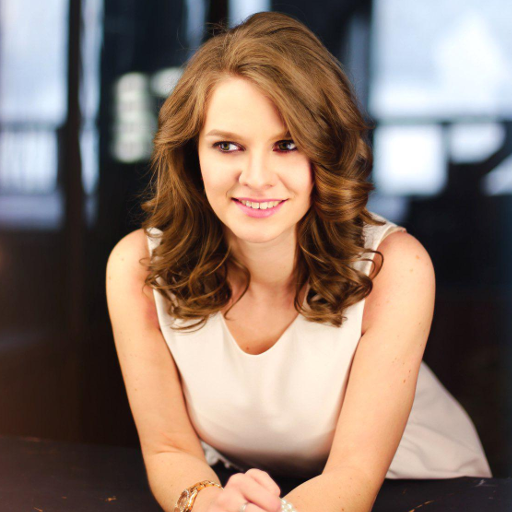}} \\
		\end{tabular}
		\caption{\label{fig:teaser}Comparative results with Deep Image Prior~\citep{ulyanov2018deep}
			on different image restoration tasks. The first example corresponds to denoising whereas 
			the second is image inpainting. 
			Our approach is able to remove the degradation and produces visually more pleasing
			results in some regions like the text and the teeth.}
\end{figure}

% 3) What do we do.
All existing methods proposed alternatives and approximations 
to the true image prior $p(\mathbf{x})$ in Equation~\ref{eq:map_problem}. 
However, with deep normalizing flows, we have an approach for a
tractable \emph{and} exact log-likelihood computation~\citep{realNVP_DinhSB17}.
Therefore, we propose to use normalizing flows for capturing the distribution of target high quality content to serve as a prior in the MAP formulation. 
In addition to this, 
the inference of the latent value that corresponds to a data point
can be done exactly without any approximation since our generative model is invertible. We use this learned bijective mapping
to express the MAP optimization process in the latent space 
and are able to obtain solutions through gradient descent. 
% 4) Evaluation and results.
In a number of experiments, we explore our approach for different degradations on data sets of varying complexity and we show that we can achieve competitive results as illustrated in Figure~\ref{fig:teaser}.

% 5) Summarize contributions. 
The contribution of this paper is three fold: 1) to the best of our knowledge, 
our work is the first using normalizing flows to learn a prior for blind
image restoration; 2) we take advantage of the bijective mapping learned
by our model to express the MAP problem of image reconstruction in latent space,
where gradient descent can be used to estimate the solution; 3) we propose
using new loss terms during model training for regularizing the latent space 
which yields a better behavior during the MAP inference.

Our paper is organized as follows. In Section~\ref{sec:background}, we recap important background regarding normalizing flow before describing
our method in Section~\ref{sec:method}. Section \ref{sec:related} covers important related work and Section~\ref{sec:experiments} discusses our experimental results. We give our conclusions in
Section~\ref{sec:conclusion}.

\section{Normalizing Flow}
\label{sec:background}
%In normalizing flow~\cite{DBLP:conf/icml/RezendeM15} the objective 
%is to map a base distribution to an arbitrary distribution to a, 
%which is done through a change of variable.
Borrowing the notation from \citet{papamakarios2019normalizing}, let's consider 
two random variables $X$ and $U$ that are related through the reversible 
transformation 
$T : \mathbb{R}^d ~\rightarrow~\mathbb{R}^d, \; \mathbf{x} = T\left(\mathbf{u}\right)$.
In this case, the distribution of the two variables are related as follows:
%--------------------------------------------------
% Equation : change of variable NF
%--------------------------------------------------
\begin{equation}
p_{X}(\mathbf{x}) = p_{U}\left(\mathbf{u}\right) \left| \det J_T\left(\mathbf{u}\right) \right|^{-1} \;,
%\;\;\; \text{where } \mathbf{u} = T^{-1}\left(\mathbf{x}\right)
\end{equation}
%-------------------------------------------------

where $\mathbf{u} = T^{-1}\left(\mathbf{x}\right)$ and $J_{T}\left(\mathbf{u}\right)$ is the Jacobian of T.
Here, the determinant preserves total probability
and can be understood as the \emph{amount} of squeezing 
and stretching of the space induced by the transformer $T$. 
The objective of normalizing flows~\citep{DBLP:conf/icml/RezendeM15}
is to map a base distribution to an arbitrary distribution 
through a change of variable. In practice, 
a series $T_1, \ldots, T_K$ of such mappings are applied
to transform the base distribution into a more complex 
multi-modal one
%--------------------------------------------------
% Equation : Normalizing flow composition
%--------------------------------------------------
%\begin{equation}
%T^{-1}\left(\mathbf{x}\right) = T^{-1}_1 \circ \ldots \circ T^{-1}_K\left(\mathbf{x}\right) = \mathbf{u}
%\end{equation}
\begin{equation}
\mathbf{x} \xleftrightarrow{T^{-1}_K} 
\mathbf{h}_{K-1} \xleftrightarrow{T^{-1}_{K-1}} 
\mathbf{h}_{K-2} \cdots
\mathbf{h}_{1} \xleftrightarrow{T^{-1}_{1}}  \mathbf{u} \;,
\end{equation}
%---------------------
\begin{equation}
\label{NF}
p_{X}\left(\mathbf{x}\right) = p_{U}\left(T^{-1}\left(\mathbf{x}\right)\right) \prod_{k=1}^{K} \left| \det \frac{d\mathbf{h}_{k-1}}{d\mathbf{h}_{k}} \right| \;,
\end{equation}
%---------------------
where we define $\mathbf{h}_K \triangleq \mathbf{x}$ and $\mathbf{h}_0 \triangleq \mathbf{u}$.
It is clear that computing the determinant of these Jacobian matrices, 
as well as the function inverses, must remain easy to allow their integration 
as part of a neural network. 
This is not the case for arbitrary Jacobians and recent successes
in normalizing flow are due to the proposition of invertible transformations 
with easy to compute determinants.

\paragraph{Normalizing flows as generative model.}
%As discussed by \citet{DBLP:conf/nips/KingmaD18}, 
%we can consider
%an image $\mathbf{x} \in \mathcal{X}$ as a high-dimensional vector
%with probability  $p\left(\mathbf{x}\right)$.
%By expressing $T_\theta$ as a sequence of bijective transformations,
Recent works~\citep{DBLP:conf/nips/KingmaD18,realNVP_DinhSB17}
have shown the great potential of using normalizing flow 
as generative model where an image observation $\mathbf{x}$ 
is generated from a latent representation $\mathbf{u}$ 
%--------------------------------------------------
% Equation : generative model
%--------------------------------------------------
\begin{equation}
\mathbf{x} = T_\theta\left(\mathbf{u}\right) \qquad \text{with} \qquad \mathbf{u} \sim p\left(\mathbf{u}\right) \;.
\end{equation}
%--------------------------------------------------
Here $\mathbf{x} \in \mathcal{X}$ is a high-dimensional vector,
$T_\theta$ denotes a composition of invertible 
transformations, and $p\left(\mathbf{u}\right)$ 
is the base distribution e.g. a normal distribution. 
Considering a discrete set $\mathcal{X}$ of $N$ natural images, 
the flow based model is trained by
%learn a parameterized distribution, $p_\theta\left(\mathbf{x}\right)$, 
by minimizing the following log-likelihood objective: 
%--------------------------------------------------
% Equation : log-likelihood 
%--------------------------------------------------
\begin{equation}
\label{eq:NF}
\mathcal{L} = \frac{1}{N} \sum_{i=1}^{N} 
	- \log p_{\theta}\left(\mathbf{x}^{(i)}\right).
\end{equation}
%corresponding to an image. 
%To learn a continuous distribution the input image has to be quantized.
%We do so by adding a sample $\mathbf{x} + \epsilon$ 
%with $\epsilon$ uniformly sampled from $\left[0, 1\right)$.
In the next section, we will describe our approach for leveraging
flow based models for various image restoration applications.

\section{Blind Image Restoration with Flow Based Priors}
\label{sec:method}
% two parts, training model, using model to enhance images.

% start with normalizing flow 
% we know how images should look like (we have distr.)
% base example (MNIST) (Distribution -> what can we do, to restore images)
% Present degratation on MNIST examples
% extending to more realisitic images (sprite)
% works well for mnist, but if we move to more complex

% section more realistic images
% define model
% define losses
% define optimization (coarse-to-fine)

%This section is separated into two main parts. 
%First we introduce the model and the optimization. 
%In the second part we describe the application of normalizing flow for image enhancement.

By training a generative flow model  as described in the previous section,
we learn a mapping $T_\theta$ from a latent space $\mathcal{U}$,
with a known base distribution $p(\mathbf{u})$, to the complex image space $\mathcal{X}$. 
In this work, we propose to use the capacity of normalizing flows to compute the exact likelihood 
of images $p_\theta(\mathbf{x})$, as prior in the image restoration problem 
%%% ---------------------
% Equation
%%% ---------------------
\begin{equation}
\label{eq:map_problem_nf}
\mathbf{x}^\star = \argmin_\mathbf{x} ~
	- \log p \left( \hat{ \mathbf{x} } | \mathbf{x} 	 \right)
	-\log p_\theta\left(\mathbf{x}\right) \;.
\end{equation}
%%% ---------------------
In addition to the prior, we also take advantage of the bijective 
mapping in normalizing flows to rewrite the optimization with respect 
to the latent $\mathbf{u}$ 
%%% ---------------------
% Equation
%%% ---------------------
\begin{equation}
\label{eq:map_problem_latents}
\mathbf{u}^\star = 
\argmin_\mathbf{u} \left[ ~ 
	- \log p \left( \hat{ \mathbf{x} } | ~ T_\theta\left(\mathbf{u}\right)\right)
	-\log p_\theta\left(T_\theta\left(\mathbf{u}\right)\right) 
	~\right] \;.
\end{equation}
%%% ---------------------
With this new formulation, we are leveraging the learned mapping
between the complex input space (the image space) and the base space
(the latent space) that follows a simpler distribution. 
This new space is more adapted for such an optimization problem.
In this work, we solve it through an iterative gradient descent, 
where each step is applied on the latents according to
%%% ---------------------
% Equation
%%% ---------------------
\begin{equation}
\label{eq:sgd}
\mathbf{u}^{t+1} = \mathbf{u}^{t} - \eta \nabla_{\mathbf{u}}~L\left(\theta,\mathbf{u}, \hat{\mathbf{x}}\right).
\end{equation}
%-------------------
Here $L\left(\theta, \mathbf{u}, \hat{\mathbf{x}}\right)$ abbreviates 
the objective defined in equation~\ref{eq:map_problem_latents}
and $\eta$ is the weighting applied to the gradient. 
We used the Adam optimizer~\citep{DBLP:journals/corr/KingmaB14} to compute 
the gradient steps. 
The model is generic and once trained on \emph{target quality} images, 
different applications can be considered by adapting the data loss term.
In this work we use a generic data fidelity term between 
the input image $\hat{\mathbf{x}}$ and the restored 
result $\mathbf{x} = T_\theta\left(\mathbf{u}\right)$:
%-------------------
\begin{equation}
\mathcal{L}_{\text{data}}\left(\hat{\mathbf{x}}, ~T_\theta\left(\mathbf{u}\right)\right) 
	= - \log p \left( \hat{ \mathbf{x} } | ~ T_\theta\left(\mathbf{u}\right)\right)
	= \mathbf{m} \odot \lambda || \hat{\mathbf{x}} - T_\theta\left(\mathbf{u}\right) ||_2 \;,
\end{equation}
%-------------------
where $\odot$ is the Hadamard product. The mask $\mathbf{m}$ 
is a binary mask that indicates pixel locations with valid color values %to be optimized, 
and allows to handle the inpainting scenario. %when the masked regions are known.
The parameter $\lambda$ controls the deviation tolerance
from the original degraded input $\hat{\textbf{x}}$. 
Next we provide details on the normalizing flow architecture used, the training losses, 
and our coarse to fine optimization procedure.

\subsection{Generative Flow Architecture}
The proposed generative model is based on the architecture 
described by \citet{DBLP:conf/nips/KingmaD18}. 
We first present the individual building layers
%These are invertible transformations with easy to compute determinants
%which facilitates their integration as part of a neural network. 
%###################################################
% Coupling layers
%###################################################
\begin{itemize}
%\paragraph{Affine transformation.}
\item\textbf{Activation normalization.} 
Proposed by \cite{DBLP:conf/nips/KingmaD18},
this is an alternative to batch normalization. 
It performs an affine transformation on the 
activations using a learned scale and bias parameter per channel.
%%%%%
\item\textbf{Invertible $1 \times 1$ convolution.}
\cite{DBLP:conf/nips/KingmaD18} 
also proposed to replace the random permutation of channels, in coupling layers 
between the transformations, with a learned invertible $1\times1$ convolution.
%%%%%
\item\textbf{Affine transformation.}
This layer is a coupling introduced 
by \citet{DBLP:journals/corr/DinhKB14}.
%Several variations exist but the core idea is 
The input is split into two partitions, 
where one is the input for the conditioner, 
a neural network to modify the channels of the second partition.
Here, the transformation is affine. 
%and the scale and bias parameters are learned 
%by a neural network.
%\paragraph{Activation normalization.} 

%	The initialization of these parameters is \textit{data dependent} 
%	and set such that the post-actnorm activations per channel 
%	for a minibatch are standard normal distributed \textit{(have zero mean and unit variance.)}.
%\paragraph{Invertible $1 \times 1$ convolution.} 
%########################################################
% Factor-out layer
%########################################################
\item\textbf{Factor-out layers.}
The objective of factoring-out parts of the base distribution~\citep{realNVP_DinhSB17}
is to allow a coarse to fine modeling by introducing conditional 
distributions and dependencies on deeper levels.
\end{itemize}

%a simplification of the representation by only further processing a part of the input features. 
%In Figure~\ref{fig:overview} we illustrate an architecture with two factor-out layers 
%where the input features are split into two parts $\mathbf{h}$ and $\mathbf{u}$.
%Formally this is expressed for one level as
%%--------------------------------------------------
%% Equation : Factor-out layer
%%--------------------------------------------------
%\begin{equation}
%(\mathbf{u}_1, \mathbf{h}_1) = T_1^{-1}\left(\mathbf{x}\right) \qquad \text{and} \qquad \mathbf{u}_0 = T_0^{-1}\left(\mathbf{h}_0\right)
%\end{equation}
%%--------------------------------------------------
%First the input $\mathbf{x}$ is mapped to $(\mathbf{u}_1, \mathbf{h}_1)$
%with the transformation $T_1^{-1}$. 
%This representation is split in two parts, with one further processed with $T_0^{-1}$. 
%The latent representation is given by $\mathbf{u} = (\mathbf{u}_0, \mathbf{u}_1)$.

\begin{figure}
	\centering
	\includegraphics[width=\columnwidth]{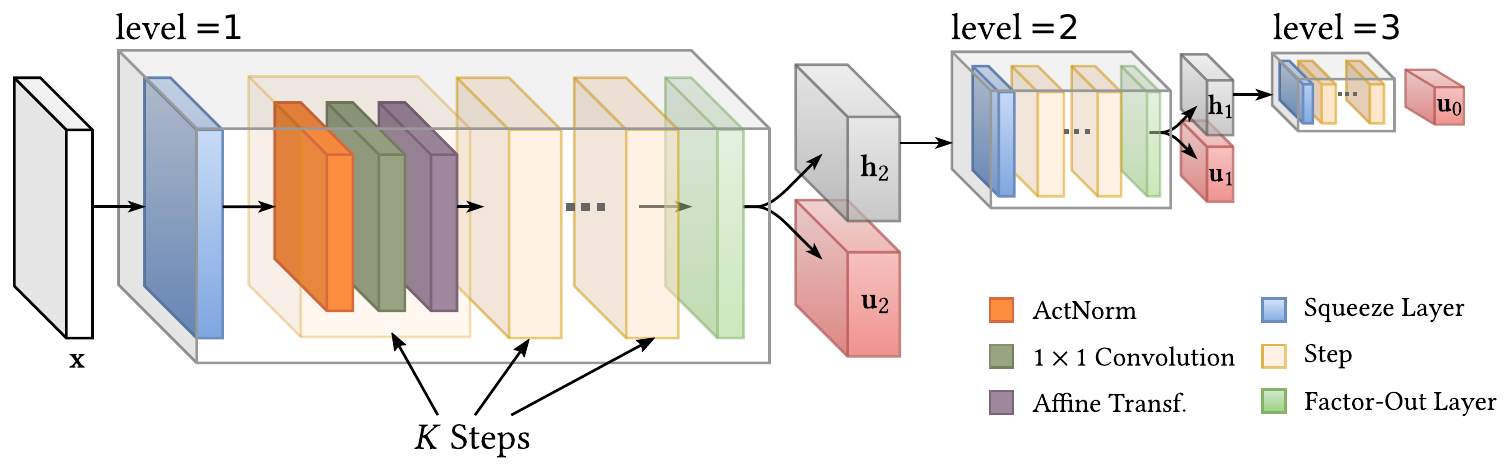}
	\caption{Overview of the normalizing flow architecture. The input image $\mathbf{x}$ 
		is processed by an $L=3$ level network, where each level consists 
		of a squeeze operation followed by a series of $K$ steps. 
		Each step is a succession of \emph{ActNorm}, \emph{$1\times1$ convolution}
		and an \emph{affine layer}. The image latent representation 
		is $(\mathbf{u_0}, \mathbf{u_1}, \mathbf{u_2})$.
		The number of levels and steps can be adapted to the complexity of the data.}
	\label{fig:overview}
\end{figure}

%\paragraph{Architecture overview.} 
Using these layers, we propose the model illustrated in Figure~\ref{fig:overview}.
It consists of $L$ levels, each one is a succession of $K$ steps, 
where a step defined as the composition of the layers:
\emph{ActNorm}, \emph{$1\times1$ convolution} and \emph{Affine}.
%The \textit{scale} and \textit{bias} parameters for the 
%affine transformation are learned by a neural network \texttt{NN}.
%A schematically illustration of a step is visualized in Figure~\ref{fig:step}.
At the end of each intermediate level, the transformed values 
(\textit{latents}) are split in two parts $\mathbf{h}_i$ and $\mathbf{u}_i$,
with the factor-out layer. 
The parameters ($\mathbf{\mu}_i, \mathbf{\sigma}_i$) of the conditional distribution 
$p\left(\mathbf{u}_i \: | \: \mathbf{h}_i\right)$ 
are predicted by a neural network. %with a ResNet with three blocks and ??? hidden channels. 
In our case, this is a zero initialized 2D convolution as proposed in~\citep{DBLP:conf/nips/KingmaD18}. 
In the experimental part and in supplementary material, we provide more 
details about the architecture used for each dataset.
%\aziz{Details of the probability model in deepest level. 
%	Are they learned too? or fixed? per position? per channel?}
%\leo{The parameters are always learned for each position. for the MNIST, we learn the $\mu$ directly as tensor (unit variance). For the sprite and and the DIV2K we learn a convolutional layer. \texttt{[mu, sigma] = split(ZeroConv2D(zeros))}, where the weights are initialized with $0$.}

%------------------------------------------
%----     Subsection : Losses + Reg.
%------------------------------------------
\subsection{Training and Latent Space Regularization}
\label{sec:regularization}
When using normalizing flows to learn a continuous distribution,
the input images have to be \emph{dequantized}. 
Following common practices in generative flows,
we redefine the negative log-likelihood objective ($nll$) 
of equation~\ref{eq:NF}
%--------------------------------------------------
% Equation : log-likelihood with noise (sim. to GLOW)
%--------------------------------------------------
\begin{equation}
\label{eq:NLL}
\mathcal{L}_{nll} = \frac{1}{N} \sum_{i=1}^{N} 
- \log p_{\theta}\left(\mathbf{x}^{(i)} + \epsilon \right) \;.
\end{equation}
%---------------
Here $\epsilon$ is uniformly sampled from $\left[0, 1\right]$. 
This model is sufficient for simple datasets as we show 
in the experimental section with the MNIST examples (see Figure \ref{fig:mnist_results}).
%\textcolor{red}{As already described in the previous section, a model trained 
%by minimizing NLL loss is not suitable for image enhancement tasks in general.. Architecture, layers, description, ...}
However for more complex data, a regularization of the learned latent 
space is needed. The main objective is to structure this space
in a beneficial way for the optimization. 
%If we train a model by minimizing the negative-log-likelihood (NLL),
%the model is not suitable for enhance complex data e.g. sprites.
%First one is to avoid any over fitting of the model to the elements in the training set,
%and avoid peaks in the learned density function 
%\textit{(at the encodings of the images $\mathbf{x} \in \mathcal{X}$)}.
%which results in the encoding of a degraded image 
%in which case the latent space becomes unsuited for 
%the gradient descent optimization.
%The second objective is to structure the latent space
%in a beneficial way for the optimization, in particular to allow
%a coarse-to-fine strategy for the optimization. 
% with 
%to be located on far from any likely latent space regions in probability
%far from distribution latents located outside 
%of this peak and the optimization diverges.
%The problem can be solved with two options: 
%(1) train with augmented (degraded) images 
%(2) introduce noise in the latent space and encode to a region instead of a single point.
%The first possibility has the downsides of specializing 
%the model to a specific degradation and to the assumption 
%that noisy data is part of the dataset 
%(this could lead to local optima in the image enhancement optimization). \\

%In Figure~\ref{fig:sprites_gaussian_5} one can see,
%that a model trained to optimize only the probability produces 
%even more degraded images due to divergence of the optimization.
%As the Sprites dataset is much more complex than the MNIST 
%dataset the optimization's starting point $\hat{\mathbf{u}}$ 
%of the optimization becomes more important. 

\paragraph{Latent-Noise loss.}
In order to enforce some regularization of the latent space, we add uniform noise 
to the latents $\mathbf{u}_\xi = \mathbf{u} + \mathbf{\xi}$ 
where $\mathbf{\xi} \sim \mathcal{U}\left(-0.5, 0.5\right)$. 
The proposed loss term
\begin{equation}
\mathcal{L}_{ln} = || T_\theta\left(\mathbf{u}_\xi\right) - \mathbf{x} ||_2
\end{equation}
penalizes parameters $\theta$ that would map back $\mathbf{u}_\xi$ far 
from the initial input image $\mathbf{x}$. 
%\textit{(at the encodings of the images $\mathbf{x} \in \mathcal{X}$)}.
%which results in the encoding of a degraded image 
It is interesting to note that this loss does not make any assumption 
regarding the degraded images, but it still results
in a latent space better suited for our optimization problem.
% \aziz{ref to fig.} \leo{(Figure~\ref{fig:sprites_gaussian_5})}.
%image it does
%on Without any prior on the degraded images, this regularizing loss is necessary 
%to maintain images $\hat{\textbf{x}}$
%as only good quality images are observed

\paragraph{Auto-Encoder loss.}
%\textcolor{red}{Motivation is different to image compression. For image compression we want a good reconstruction, while for image enhancement we want a good starting point}
If we consider the model illustrated in Figure~\ref{fig:overview},
the image $\textbf{x}$ is mapped to its representation 
$(\textbf{u}_0,  \textbf{u}_1, \textbf{u}_2)$.
From only the latent value $\mathbf{u}_0$,  
we compute $\tilde{\mathbf{x}}$ by sampling the most likely
intermediate values $\tilde{\mathbf{u}}_l \sim p\left(\mathbf{u}_{l}\:|\:\mathbf{h}_{l}\right)$.
Since we use a Gaussian distribution, this corresponds to the mean value 
of the predicted distribution. 
% \leo{should we explicitly mention, that we use the mean value of this distribution? / should we use here the $\mathbf{u}_\epsilon$ instead?}
The proposed loss
\begin{equation}
\mathcal{L}_{ae} = || \tilde{\mathbf{x}} - \mathbf{x} ||_2
\end{equation}
%and $\textbf{u}_2$. most probable 
%encodings for the intermediate levels. 
%
%by minimizing the mean squared error between the input image $\mathbf{x}$
%and the image we obtain when decoding the latents of the deepest level $\mathbf{x}'$.\\
%To compute $\mathbf{x}'$ we decode the bottleneck $\mathbf{u}_0$ and 
%the most probable encodings for the intermediate levels. 
%The full decoded image relies only on the information saved in the bottleneck.
%The intermediate encodings $\mathbf{u}'_l$ are obtained by 
%taking the most likely sample from the learned distributions \qquad\qquad
%$\mathbf{u}'_l \sim p\left(\mathbf{u}_{l}\:|\:\mathbf{h}_{l}\right)$. 
%
forces the model to store sufficient information in the deepest level
to reconstruct the image. % (e.g. $\mathbf{u}_0$ in Figure \ref{fig:overview}). 
This allows a more robust coarse-to-fine strategy during the optimization. 

The final training loss for the normalizing flows is
\begin{equation}
\mathcal{L} = \mathcal{L}_{nll} + \beta_{ln} \mathcal{L}_{ln} + \beta_{ae} \mathcal{L}_{ae},
\end{equation}
where $\beta_{ln}$ and $\beta_{ae}$ are the weightings for each loss term.
We used $\beta_{ln} = 100$ and $\beta_{ae} = 1$. The ablation study
in the experimental section shows the necessity of training 
the generative flow model with all these loss terms.

\subsection{Coarse-To-Fine Optimization}
\label{sec:coarse-to-fine}
%In order to find the best enhanced reconstruction, 
%we search in proximity of the encoding of the degraded image. 
%Since the latent space itself could be highly non-linear, 
%the optimization could end up in a local minimum.
%The introduced multi-scale architecture reduces the size of the latent space 
%by half at each level. Through the introduced loss term, we force
%the model to store coarse information about the image in the lower levels
%while the detailed information will be encoded in the higher levels.
%In order to overcome this problem, we exploit the architecture
%of the flow by optimizing each level separately. This allows
%the optimization procedure to focus only on a fraction of the encoding.
%More over, the closer we are on a proper encoding of $\mathbf{x} \in \mathcal{X}$
%in the lower levels, the better are the estimates for the parameters of higher levels.
%Therefore, high probable latent vector in the lower 
%levels lead to better estimates in the higher levels. 
The optimization procedure described in Equation~\ref{eq:sgd}
is iterative and we need to set its initial value $\textbf{u}^0$.
In order to choose a good starting point, we leverage the introduced 
multi-scale architecture.  
Our starting point is
\begin{equation}
\textbf{u}^0 = (\hat{\textbf{u}}_0, \tilde{\textbf{u}}_1, \tilde{\textbf{u}}_2) 
	\quad \text{with} \quad 
	\hat{\textbf{u}}_0 \quad \text{defined by} \quad T^{-1}_\theta(\hat{\mathbf{x}}).
\end{equation}
The values of the other components, $\tilde{\textbf{u}}_1$ and $\tilde{\textbf{u}}_2$, are 
sampled as the mean values of the respective predicted distributions.
$p\left(\mathbf{u}_{1}\:|\:\mathbf{h}_{1}\right)$ and $p\left(\mathbf{u}_{2}\:|\:\mathbf{h}_{2}\right)$.
As our auto-encoder loss enforces the possibility to reconstruct the image
from $\hat{\textbf{u}}_0$ only, this lowest level contains coarse image information
while details are stored in the upper levels. 
This is advantageous for image restoration tasks where the degradation
often affects the \emph{detail} of an image. 

Given this starting point, the optimization is done in a coarse-to-fine fashion. 
First, only the lowest level variables are optimized while the upper levels 
are respectively sampled from the predicted means. 
These are then progressively included in the optimization
%%% ---------------------
% Equation
%%% ---------------------
\begin{align}
\mathbf{u}_0^{t+1} &= \mathbf{u}_0^{t} - \eta \nabla_{\mathbf{u}_0}~L\left(\theta,\mathbf{u}, \hat{\mathbf{x}}\right) \;, \\
(\mathbf{u}_0,\mathbf{u}_1)^{t+1} &= (\mathbf{u}_0,\mathbf{u}_1)^{t} - \eta \nabla_{(\mathbf{u}_0,\mathbf{u}_1)}~L\left(\theta,\mathbf{u}, \hat{\mathbf{x}}\right) \;, \\
(\mathbf{u}_0,\mathbf{u}_1,\mathbf{u}_2)^{t+1} &= (\mathbf{u}_0,\mathbf{u}_1,\mathbf{u}_2)^{t} - \eta \nabla_{(\mathbf{u}_0,\mathbf{u}_1,\mathbf{u}_2)}~L\left(\theta,\mathbf{u}, \hat{\mathbf{x}}\right) \;.
\end{align}
%-------------------
With this coarse-to-fine scheme, we are able to incrementally refine
the reconstructed images by making sure that 
the lower level information is correct first.

\section{Related Work and Discussion}
\label{sec:related}
Despite the success of supervised deep learning approaches for dedicated image restoration problems such as 
super-resolution~\citep{wang2018fully,zhang2018image}, 
denoising~\citep{zhang2017beyond}, 
inpainting~\citep{pathak2016context} or
a combination of them~\citep{park2017joint},
one important drawback
is the need for retraining whenever the specific degradation or its parameters change.
Some recent works~\citep{CDW19blindsr,bell2019blind} have investigated 
the blind setting for super-resolution. However that concerns the parameters
of the degradation only and such solutions are not applicable to an unknown degradation.

When addressing the blind restoration problem, the common approach
is to consider the Bayesian perspective where recovering the original 
image is expressed as solving a maximum a posteriori
(MAP) problem. The objective function consists of a fidelity term
and a regularization term. The fidelity term can be problem specific
and easier to express than the prior that is supposed to reflect 
desired properties of the reconstructed image. 
Existing handcrafted priors are based on total variation~\citep{TVrudin1992},
gradient sparsity~\citep{sparsityfergus2006} or the dark pixel prior~\citep{he2010single}.
Recently, several works have investigated the usage of CNNs as priors. 
For example, \citep{rick2017one,zhang2017learning} show how a deep CNN trained for image denoising can effectively 
be used as prior in various image restoration tasks. 
Additionally, \cite{meinhardt2017learning} 
provide new insights on how the denoising strength of the neural network
relates to the weight on the data fidelity term.
\cite{bigdeli2017deep} define a utility function that includes the
smoothed natural image distribution and relate this 
to denoising autoencoders.  
In a different direction, \cite{ulyanov2018deep} 
showed that an important part of the image statistics is already 
captured by the structure of a convolutional image generator itself, 
independent of any learning. This work was further analyzed 
from a Bayesian perspective~\citep{cheng2019bayesian}
and combined with a denoising autoencoder prior~\citep{mataev2019deepred}.

The idea presented in our work stems from recent developments 
in normalizing flows~\citep{DBLP:journals/corr/DinhKB14,realNVP_DinhSB17,DBLP:conf/nips/KingmaD18}
and their promising capacity of learning a bijective mapping from
a space with a prescribed distribution to the complex space of images,
additionally providing exact log-likelihood tractability.
Using a learned prior that only depends on properties of high quality
images is an exciting direction, as this removes the need to rely
on other assumptions that are either explicit, in the case of handcrafted solutions,
or implicit in the case of denoising autoencoders.
This work is a first step demonstrating the potential of normalizing
flows in image restoration tasks. 
%-> basically we're trying to say that noise flow is an interesting first step and so are we? maybe we can remove this? or how important is it to mention noise flow here?} %\aziz{are next sentences too negative?} 
%As such it does not yet outperform other priors and supervised approaches,
We believe this is an exciting new direction that furthermore is expected 
to benefit from improvements and research that generally explores normalizing flow as a generative model.

\section{Experiments}
\label{sec:experiments}
\begin{figure}
	\centering	
	\resizebox{\columnwidth}{!}{
		\begin{tabular}{ccccccccccc}
			& 
			$N(0,20)$ & 
			$N(0,30)$ & 
			$N(0,50)$ & 
			JPEG 30 & 
			JPEG 10 & 
			JPEG 5 & 
			$U(\pm20)$ & 
			$U(\pm40)$ & 
			Mask(10) & 
			\makecell[l]{$U(\pm40)$ $\circ$\\JPEG 10 $\circ$\\Mask(10)}\\
			
			\rotatebox[origin=l]{90}{\quad GT} &
			\includegraphics[scale=1.5]{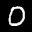} &
			\includegraphics[scale=1.5]{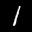} &
			\includegraphics[scale=1.5]{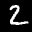} &
			\includegraphics[scale=1.5]{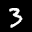} &
			\includegraphics[scale=1.5]{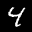} &
			\includegraphics[scale=1.5]{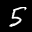} &
			\includegraphics[scale=1.5]{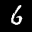} &
			\includegraphics[scale=1.5]{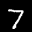} &
			\includegraphics[scale=1.5]{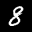} &
			\includegraphics[scale=1.5]{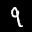} \\
			
			\rotatebox[origin=l]{90}{\quad In} &
			\includegraphics[scale=1.5]{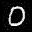} &
			\includegraphics[scale=1.5]{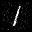} &
			\includegraphics[scale=1.5]{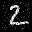} &
			\includegraphics[scale=1.5]{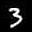} &
			\includegraphics[scale=1.5]{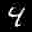} &
			\includegraphics[scale=1.5]{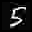} &
			\includegraphics[scale=1.5]{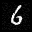} &
			\includegraphics[scale=1.5]{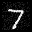} &
			\includegraphics[scale=1.5]{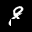} &
			\includegraphics[scale=1.5]{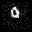} \\
			
			\rotatebox[origin=l]{90}{\quad Out} &
			\includegraphics[scale=1.5]{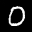} &
			\includegraphics[scale=1.5]{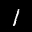} &
			\includegraphics[scale=1.5]{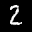} &
			\includegraphics[scale=1.5]{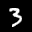} &
			\includegraphics[scale=1.5]{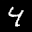} &
			\includegraphics[scale=1.5]{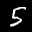} &
			\includegraphics[scale=1.5]{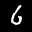} &
			\includegraphics[scale=1.5]{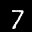} &
			\includegraphics[scale=1.5]{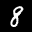} &
			\includegraphics[scale=1.5]{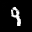} \\
		\end{tabular}
	}
	\caption{Results produced by a single-level normalizing flow trained on the MNIST dataset. Each column corresponds to a different type of degradation. From top to bottom the ground truth, the degraded image and the reconstructed image are shown. %(see text for details about the degradation types).
		%The digites in the first three columns are degraded with random noise $N(0, \sigma)$. JPEG compression with different quality settings is used for digits three to five. Additive uniform noise $U(\pm A)$ was added to digits six and seven. A patch of size 10x10 was masked out of digit eight. Finally, for digit nine, we applied a composition of all.
	}
	%All reconstructions were obtained using the parameters $lr=0.05$, $\lambda=999$, $n\_iterations=300$.
	\label{fig:mnist_results}
\end{figure}

In this	section we explore the usage of our proposed solution
for different blind image restoration tasks. 
%we first describe the possible degradations 
%we investigate in this work. 
We show results on two synthetic datasets, the MNIST and 
the self generated Sprites, and on real images. We also include comparisons 
with the Deep Image Prior~(DIP)~\citep{ulyanov2018deep}.

Since we do not focus on a specific degradation during training, 
our proposed approach can be applied on various types of restoration 
problems. In this work we present results on $3$ different 
types of image degradation: noise (uniform and normal), 
JPEG compression artifacts, and missing regions. 
% and down scaled images
%as well as a composition of multiple degradations.\\
The noisy images are generated by adding i.i.d. samples of
noise to the pixel values, with noise distributed according to $\mathcal{U}\left(\text{min}, \text{max}\right)$ 
or $\mathcal{N}\left(0, \sigma\right)$ . 
%Note, that the pixel values are not clipped and can potentially exceed color values.
The varying degrees of JPEG artifacts are generated by using 
different levels (10 to 70) for the JPEG compression.
% \footnote{as implemented in Pillow (\url{https://pillow.readthedocs.io/en/stable/}).}.
For the inpainting task, we masked multiple regions of size $10\times10$ pixels.
%\textcolor{red}{mention pre-initialization: for the MNIST dataset, 
%	the missing values were filled with 0. 
%	For the RGB datasets (sprite / DIV2k) the inpainting algorithm proposed in \citep{DBLP:journals/jgtools/Telea04} was used 
%	to get an initial prediction for the missing region. 
%	This step is crucial, since the encoding of 
%	the degraded image is used as the starting point
%	for the optimization algorithm.
%A inappropriate initial estimate can lead to a the starting outside of the support of the learned distribution. Since the starting point would be located in a flat area (low probability) with gradients close to zero, the optimization would be slowly converging or potentially diverging.}
An overview of the used degradations is visualized in Figure~\ref{fig:mnist_results}. 

\paragraph{MNIST results.} As a first step we tested our flow based 
image prior on the well studied MNIST dataset \citep{lecun1998gradient}. 
Given the simplicity of this dataset, the model used for this experiment
consists of a single-level $L = 1$ with $K = 16$ steps.
We choose the base distribution $p\left(\mathbf{u}\right)$ to be a Gaussian 
with unit variance and a trainable mean. 
Further, a ResNet~\citep{he2016deep} with $2$ blocks and $C = 128$ 
intermediate channels, was used to learn the parameters for 
the affine transformations.

Given a degraded image $\hat{ \mathbf{x} }$ the goal is to find the 
most likely image $\mathbf{x}^\star$ by solving the 
optimization problem of Equation~\ref{eq:map_problem_latents}.
Given the simplicity of the data set, we use the mean 
of the base distribution $p\left(\mathbf{u}\right)$
as starting point $\mathbf{u}^{0}$.
It can be seen in Figure~\ref{fig:mnist_results} that
this is sufficient to enhance the binary digits for any degradation. 
%\leo{(somehow emphasize that MNIST is easy to enhance, since its an easy dataset)}.
A related experiment was conducted by \cite{DBLP:journals/corr/DinhKB14},
where the degraded digits were enhanced by maximizing 
the probability of the image trough back propagation to the pixel values. 
This is equivalent to only considering the prior 
term in Equation~\ref{eq:map_problem_latents}. 
%\aziz{Should we say more about \cite{DBLP:journals/corr/DinhKB14}?} \leo{I think thats good}

\paragraph{Sprites results.} To handle this larger and more complex data set, 
we increased the capacity of our flow based prior.
% The model is based on the architecture proposed by 
% \cite{DBLP:conf/nips/KingmaD18} but with less steps and different transformation networks. 
We use $L = 3$ levels, with $K = 8$ steps each. 
In the optimization, the learning rate $\eta$ and the data weighting term $\lambda$ 
are set to $1$ and $99$, respectively. 
The gradient descent is done in a coarse to fine way 
(see section~\ref{sec:coarse-to-fine}), %(starting at $\hat{\mathbf{u}}_0$)
each time with $50$ update steps per level before including the next one.
When all latent levels are included, an additional $150$ optimization steps are performed.
% to optimize the full model
%\leo{.. to optimize the full model?}
%optimizing all levels. The total computation time 
%is approximately $50$ sec. per image.
%All the distributions are normal with parameters learned 
%by neural networks: the conditional factor-out distributions 
%are learned by a single 2D convolution (3x3 kernel), and
%the parameters for the base distributions are learned by 
%the zero initialized 2D convolution introduced by~\cite{DBLP:conf/nips/KingmaD18}.

\begin{figure}
	\centering
	%	\hspace{-0.75cm}
	\resizebox{\columnwidth}{!}{
		\begin{tabular}{cccccc}
			Ground Truth & Input & 
			$\mathcal{L}_{nll}$ & 
			$\mathcal{L}_{nll} + \mathcal{L}_{ln}$ & 
			$\mathcal{L}_{nll} + \mathcal{L}_{ae}$ & 
			All \\
			
			\includegraphics{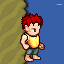} &
			\includegraphics{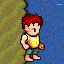} &
			\includegraphics{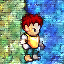} &
			\includegraphics{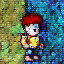} &
			\includegraphics{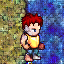} &
			\includegraphics{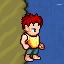} \\
			
			\includegraphics{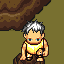} &
			\includegraphics{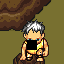} &
			\includegraphics{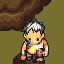} &
			\includegraphics{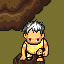} &
			\includegraphics{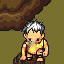} &
			\includegraphics{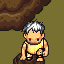} \\
			
			\includegraphics{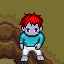} &
			\includegraphics{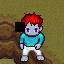} &
			\includegraphics{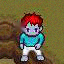} &
			\includegraphics{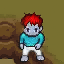} &
			\includegraphics{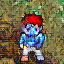} &
			\includegraphics{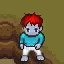} \\
		\end{tabular}
	}
	\caption{Restoration of degraded Sprites: first row corresponds to a Gaussian noise 
		with $\sigma=5$, second row is inpainting and last row combines denoising, inpainting and 
		JPEG artifact removal. Columns correspond to different normalizing flow models, each one
		trained with the indicated loss term. Results show the importance of using all the proposed 
		loss terms (see text for details).}
	\label{fig:sprites}
	\vspace{-0.5cm}
\end{figure}

Figure \ref{fig:sprites} shows image restoration results on this data set:
The first row corresponds to a denoising task, the second is image inpainting
and the last combines both in addition to compression artifact removal. 
Note that these images were not observed during training.
As the data becomes more complex, we can see the importance 
of the regularization losses proposed in Section~\ref{sec:regularization}.
Using the negative-log-likelihood loss ($\mathcal{L}_{nll}$) is clearly 
not sufficient, and a prior trained only with this term is not suited 
for the latent space optimization. The most important improvement comes from 
using the latent-noise loss ($\mathcal{L}_{ln}$).
This regularization enforces neighboring elements in latent space 
to be mapped back to similar images. 
This is highly beneficial to the gradient descent procedure in latent space
and a prior trained with this loss already leads to some good restoration results.
Finally, a coarse-to-fine approach is able to handle most cases,
in particular high intensity noise levels. This requires training 
the normalizing flow model with the additional auto-encoder loss ($\mathcal{L}_{ae}$).
\begin{figure}
	\hspace{-0.2cm}
%	\centering
	\includegraphics[width=1.02\columnwidth]{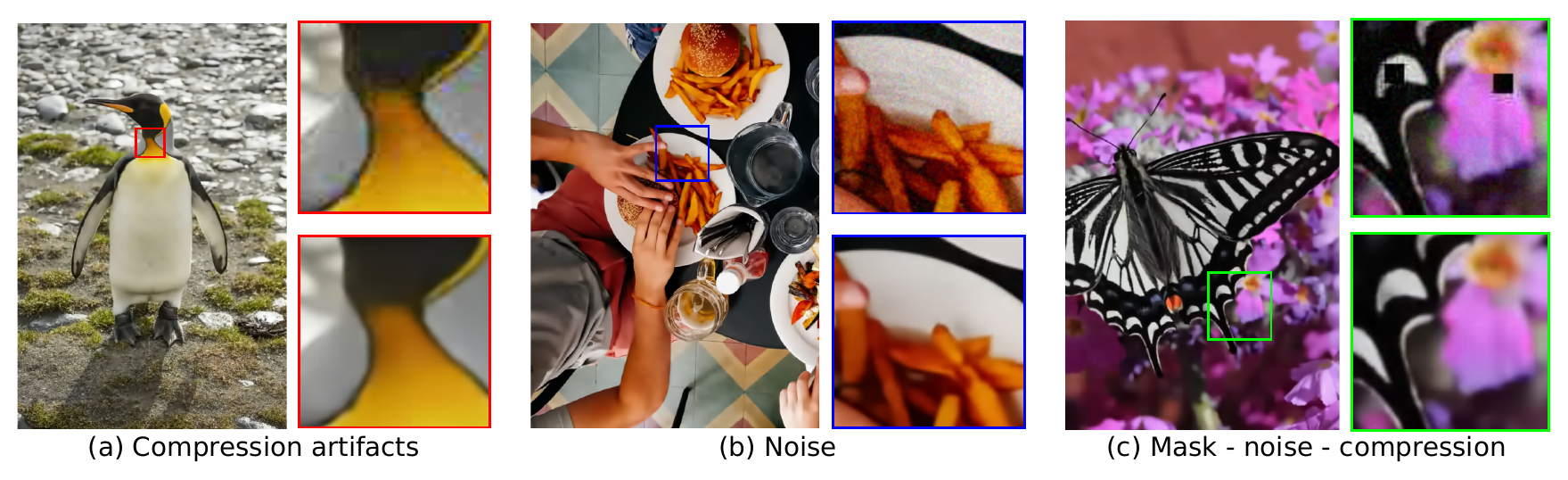}
	\caption{Results on DIV2K dataset. The proposed prior is used to restore 
		arbitrary size images. Degradations include: (a) JPEG compression artifacts;
		(b) denoising; and (c) a combination of masked regions, noise and compression
		artifacts. }
	\label{fig:div2k}
\end{figure}

\paragraph{Blind image restoration.}
We show that the proposed model is applicable to the restoration of generic images. 
In order to do so, the model must generalize 
to patches of high resolution good quality images.
For this we use the DIV2K dataset~\citep{DBLP:conf/cvpr/AgustssonT17} 
that serves as training and test set for most image super-resolution works. 
We use the same train/test split with $800$ images in the training set 
and $100$ in the test set. Training is done %for \aziz{X epochs?} 
on random image patches of size $64\times64$.
The normalizing flow architecture used here is very similar to the 
one described for the Sprites (see supplementary material for details).
%with the following changes: 
%the number of steps is set to $K = 4$ at each level; 
%the number of hidden dimensions is increased to $c = 256$ in the ResNets;
%Finally, \aziz{not sure to understand the next thing:}
%we increased the receptive field in the conditional distributions 
%(5 convolutional layers) and added a single dropout layer...
The restoration of full images of arbitrary size can be done by 
reconstructing each patch individually. A margin is used to avoid 
boundary artifacts between patches.
%More specifically for patches of $64\times64$ pixels
%we use a margin $M=4$ pixels. Neighboring patches overlap in a
%region of width $2M$ (see supplementary material for illustration). 
%This overlap between adjacent patches yields more consistent 
%results in boundary regions.
Restoration results are presented in Figure~\ref{fig:div2k}
for different image degradations. For each example,
we show the full resolution result, then focus
on a part of the image, illustrating the change.

\begin{figure}[t]
	\begin{minipage}{0.52\textwidth}
		\hspace{-0.25cm}
		\includegraphics[width=1.12\columnwidth]{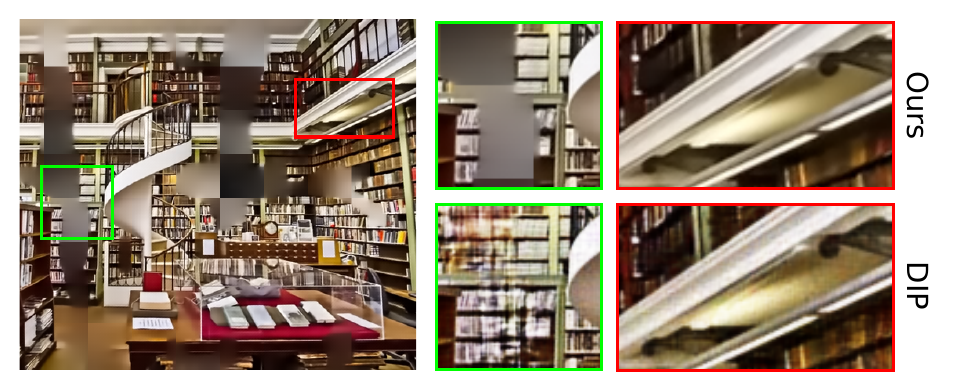}
		\caption{\label{fig:dip} Compared to DIP, restoring large missing regions
		is not possible (green), but on this example it produced better 
		denoising results (red).}
	\end{minipage}
	\hspace{0.5cm}
	\begin{minipage}{.42\textwidth}
		\vspace{0.35cm}
		\begin{tabular}{l r r }
			\midrule
			\textbf{Type of degradation} & \textbf{DIP} &  \textbf{Ours} \\ 
			\midrule
			JPEG artifacts & $27.91$ & $\mathbf{30.29}$ \\
			Noise & $\mathbf{29.45}$ & $28.99$ \\
			Multiple degradations & $25.96$ & $\mathbf{29.87}$\\
			\midrule
			\vspace{0.25cm}
		\end{tabular} 
		\caption{\label{fig:quantitative_eval}Quantitative evaluation on DIV2K using PSNR (see text for details).}
	\end{minipage}%
\end{figure}
%--------------------------------
% Table of quant eval with other SR Methods 
%--------------------------------
%\begin{figure}[t]
%	
%
%
%%\resizebox{\columnwidth}{!}{
%\begin{wrapfigure}{r}{0.4\textwidth}
%
%	\vspace{-0.5cm}
%	\hspace{-0.3cm}
%	
%\end{wrapfigure}
%%	}
%----------------------

\vspace{-0.25cm}
\paragraph{Comparison with Deep Image Prior (DIP).}
We first compare the two methods on the images presented in the original 
DIP paper~\citep{ulyanov2018deep}.
We use our same model trained on the DIV2K dataset. 
%Results are presented in and Figure~\ref{fig:dip}.
We show competitive restoration results~(Figure~\ref{fig:teaser}),
producing even visually more pleasing reconstruction than DIP 
on some regions (such as the text and the mouth). 
The main limit in our case is the patch size used during training. 
Because of this, it is not possible to inpaint 
large masked regions such as in the library image~(Figure~\ref{fig:dip}).
Interestingly however, in this case background regions are better denoised.
We also conduct a quantitative evaluation with results presented in
Figure~\ref{fig:quantitative_eval}. Using the test set from DIV2K,
we try to restore different degradations: Noise ($\mathcal{N}(0,5)$), 
JPEG artifacts and a combination of artifact removal, denoising and inpainting. 
For this comparison it is unclear how to best set the number of iterations
for the DIP. To handle this, we started from the observation that our method
converges to the result in approximately $1$ hour of computation. 
Using the DIP online implementation, this corresponds to around $10$k 
optimization steps on the denoising task. We used this maximum number 
of steps as the threshold for all images and degradations of the test set.
The evaluation using PSNR as error metric (Figure~\ref{fig:quantitative_eval}),
demonstrates that our approach is able to achieve competitive results and even outperform 
DIP on some of the restoration tasks.

\section{Conclusion}
\label{sec:conclusion}

In this paper, we explored using normalizing flows for capturing 
the distribution of target high quality content 
to serve as a prior in blind image restoration. 
To the best of our knowledge, this is the first time such a direction 
is explored. One advantage of this formulation is the learned bijective mapping
from image to latent space that we use to express 
the MAP problem of image reconstruction in latent space.
%where gradient descent can be used to estimate the solution. 
We also show the importance of using regularizing losses during 
training. Finally, we present experimental results 
illustrating the capacity of the proposed solution
to handle different degradations on data sets of varying complexity.
We believe this is an exciting new direction as there is still a lot
of potential for improvement.
% as new research works 
% explore new bijective layers, architecture designs and refined training procedures.

%\section*{Broader Impact}
%The proposed approach can have an impact on image generation 
%and image enhancement. As a result it shares the same broader impact: 
%It can help restore old content like old photo prints,
%heritage footage and classic films to be shared with
%new audiences. On the negative side, it also means 
%potential new ways to create images for misinformation purposes. 

\bibliography{imcodec_nf}
\bibliographystyle{icml2020}

%\maketitle
\newpage
\appendix

\section{Supplementary Material}

\subsection{Additional Comparison with Deep Image Prior}
We provide an additional comparison with Deep Image Prior
for the task of compression artifact removal.

\begin{figure}[h]
	\hspace{-0.5cm}
	\includegraphics[width=1.1\textwidth]{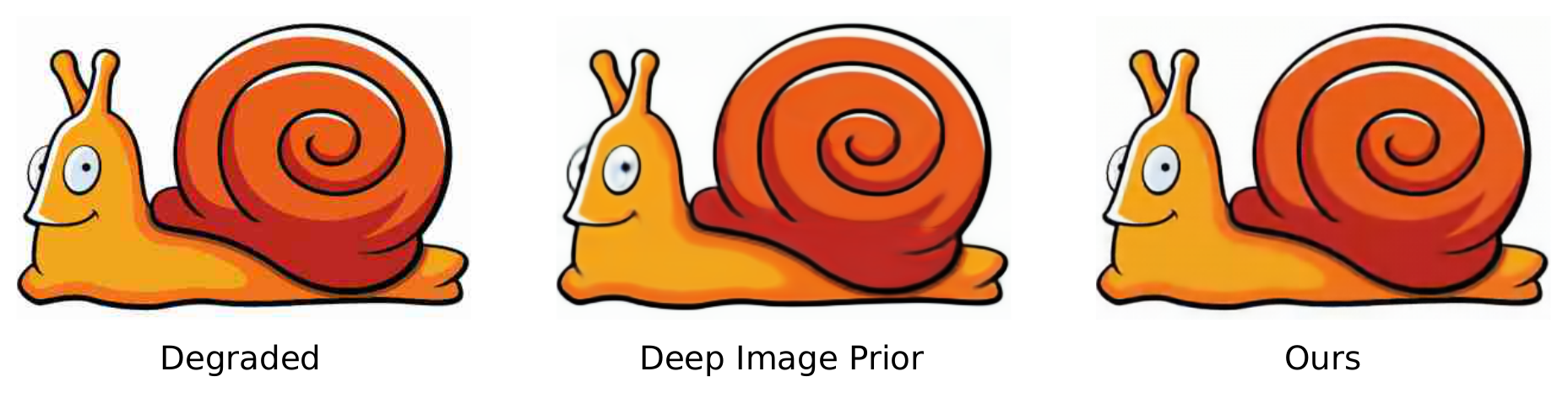}
	\caption{JPG artifact removal. We can observe that our results are sharper around the eyes.}
	\label{fig:snail}
\end{figure}

\subsection{MNIST}
%\paragraph{Architecture.}
\label{par:mnist_architecture}
For MNIST the network architecture is kept simple, only consisting of a single level. 
We use $K=16$ steps in our model. Due to the fact that squeezing layers require 
the input's height and width to be divisible by two the input 
images are zero-padded to size $32\times32$. % As 32 is a power of two this made experimentation with different architectures easy.

\begin{figure}[h]
	\centering
	\includegraphics[width=\textwidth]{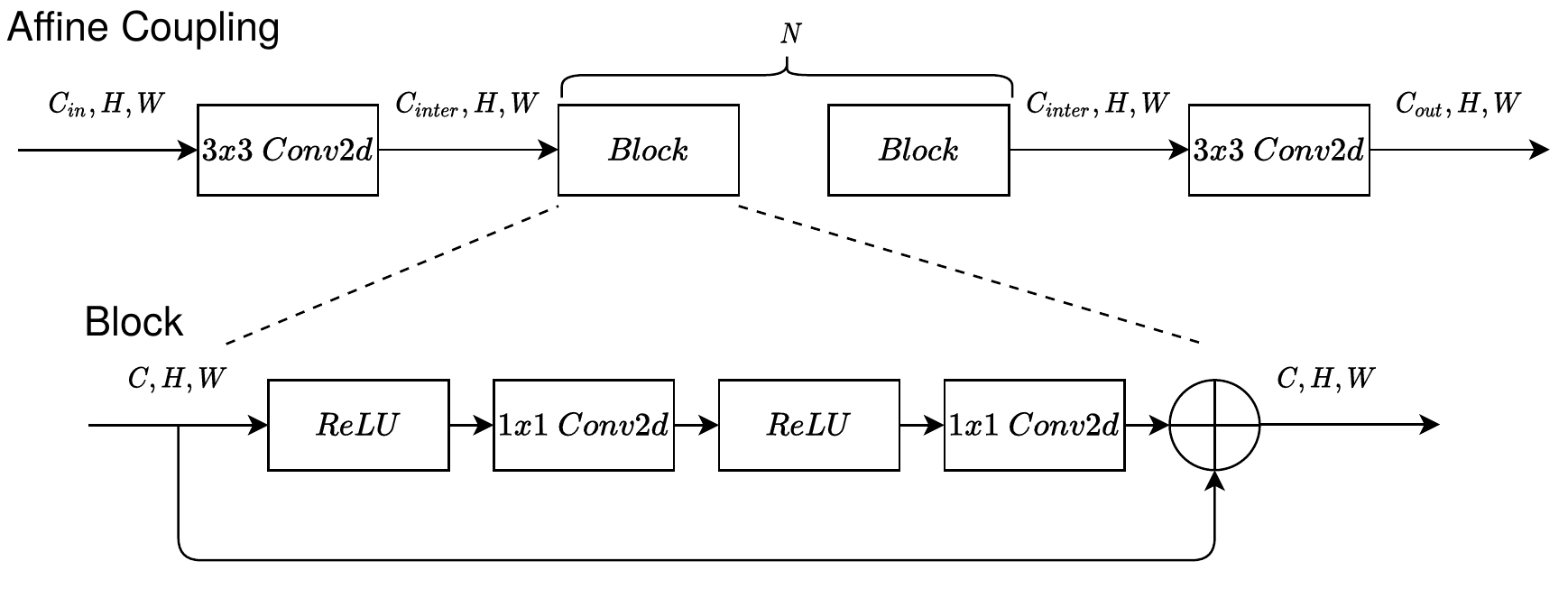}
	\caption{Details of the affine coupling transform. $3x3\ Conv2d$  and $1x1\ Conv2d$ refer to standard 2D convolutions using a kernel size of 3x3 and 1x1 respectively. The $"+"$ at the end of the block is an element wise addition.}
	\label{fig:coupling_transform_architecture}
\end{figure}

As coupling transform we use the one depicted in Figure~\ref{fig:coupling_transform_architecture} with two blocks ($N=2$) and $128$ intermediate channels ($C_{inter} = 128$). Finally, we choose a Gaussian with unit variance as our base distribution. The Gaussian's mean is set to a trainable parameter. All other parameters are listed in Table~\ref{table:mnist_training_params}. 

\begin{table}[h]
	\centering
	\begin{tabular}{l l}
		\textbf{Parameter} & \textbf{Value} \\
		\hline
		\# levels & 1\\
		\# flow blocks per level ${N_f}$ & 16\\
		Affine coupling $C_{inter}$ & 128\\
		Base distribution $p(\mathbf{u}_0)$ & $\mathcal{N}(\mu, 1)$\\
		optimizer & Adam\\
		learning rate & $10^{-4}$ \\
		batch size & 50 \\
		\# steps & $10^5$ \\
		max gradient value & $10^5$\\
		max gradient $L_2$-norm & $10^4$\\
		\hline
	\end{tabular}
	\caption{Details of architecture and training for the MNIST experiments}
	\label{table:mnist_training_params}
\end{table}

\subsection{Sprites}
Each image in the Sprites dataset consists of a figure performing some pose in front of a random background. Figures are centered in the image and have varying color for hair and clothing. Each image is of size 64x64. 
Dataset will be made available upon acceptance. 

\paragraph{Architecture.}
For this experiment, the number of levels is set to $L=3$ and each level has $K=8$ steps.
The distributions $p(\mathbf{u}_1|\mathbf{h}_1)$ and $p(\mathbf{u}_2|\mathbf{h}_2)$ depend on a function which computes mean $\mu(\mathbf{h}_i)$ and variance $\sigma(\mathbf{h}_i)$. 
We call this function the context encoder. 
A single 2D convolution with kernel size $3\times3$ and twice the number of output dimension as input dimensions is used as the context encoder. The context encoder's output is then split in half along the channel dimension. One half is used as $\mu(\mathbf{h}_i)$, the other as $\sigma(\mathbf{h}_i)$. The convolutions weight and bias are initialized to zero for stability reasons.
The other parameters for the Sprites dataset are listed in Table~\ref{table:sprites_training_params}.

\begin{table}[h]
	\centering
	\begin{tabular}{l l}
		\textbf{Parameter} & \textbf{Value} \\
		\hline
		\# levels ($L$) & 3\\
		\# flow steps per level (${K}$) & 8\\
		Affine coupling  $C_{inter}$ & 128\\
		Base distribution $p(\mathbf{u}_1|\mathbf{h}_1)$, $p(\mathbf{u}_2|\mathbf{h}_2)$ & $\mathcal{N}(\mu(\mathbf{h}_i),\ Diag(\sigma(\mathbf{h}_i)))$\\
		Base distribution $p(\mathbf{u}_0)$ & $\mathcal{N}(\mu,\ Diag(\sigma))$\\
		Context Encoder $p(\mathbf{u}_1|\mathbf{h}_1)$, $p(\mathbf{u}_2|\mathbf{h}_2)$ & zero initialized 2D Convolution, kernel size 3x3\\
		optimizer & Adam\\
		learning rate & $10^{-4}$ \\
		batch size & 20 \\
		\# steps & $10^5$ \\
		max gradient value & $10^5$\\
		max gradient $L_2$-norm & $10^4$\\
		latent noise magnitude &  $\pm 0.5$ \\
		latent noise loss ($\beta_{ln}$) & 100 \\
		autoencoder loss ($\beta_{ae}$)& 1 \\
		\hline
	\end{tabular}
	\caption[Sprites training specification]{Sprites training specification.}
	\label{table:sprites_training_params}
\end{table}

\subsection{DIV2K}
\begin{figure}[h]
	\centering
	\includegraphics[width=\textwidth]{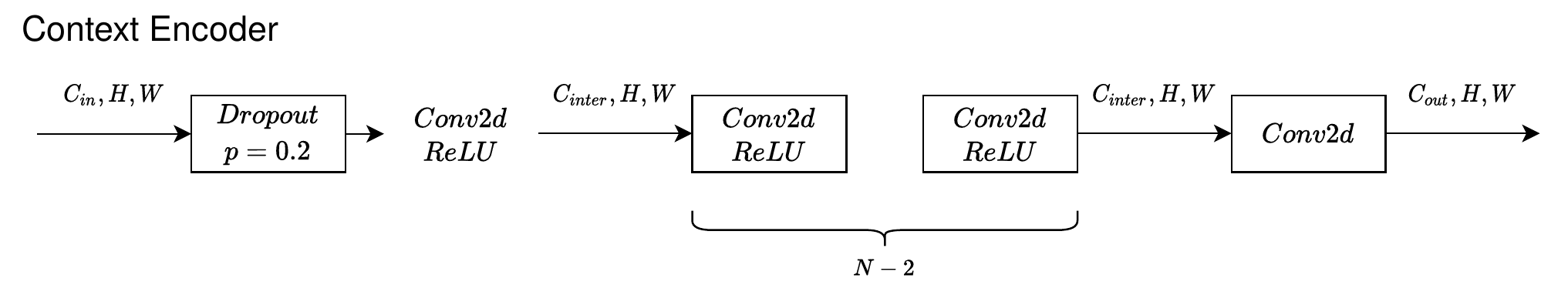}
	\caption[DIV2K context encoder architecture]{Architecture of the context encoder used for the DIV2K example. A dropout layer with $p=0.2$ is used as the first layer to prevent overfitting. The last convolution's weight and bias are initialized to zero for stability reasons.}
	\label{fig:context_encoder_architecture}
\end{figure}

%\paragraph{Architecture.}
The number of levels in the architecture is set to $L=8$ with $K=4$ steps per level. 
The number of intermediate channels in the coupling transforms is $256$. 
The context encoder architecture is deepened from $1$ to $5$ convolutional 
layer as is illustrated in Figure~\ref{fig:context_encoder_architecture}
and a dropout layer is added to the beginning. 
All the architecture parameters are listed in Table~\ref{table:div2k_training_params}.

In addition to this we found that at test time the optimization was faster when
the model was trained with additional noise on the images.

%The image noise loss works similarly to the latent noise loss, except that the noise 
%is added to the image $\textbf{x}$, and distortion is measured in the image encoding $\textbf{u} = T^{-1}_\theta (\textbf{x})$.
%%$\eta$ is uniformly sampled in $\pm 10$
%
%\begin{equation}
%\beta_{In} ||T^{-1}_\theta (\textbf{x}) - T^{-1}_\theta(\textbf{x}+\eta)||_2

The \textit{Image-Noise}-loss $\mathcal{L}_{in}$ works similarly to the \textit{Latent-Noise}-loss $\mathcal{L}_{ln}$ (see Equation 13 in the main paper) except the noise is added to the image $\mathbf{x}$ and distortion is measured on the encoding $\mathbf{u} = T^{-1}_\theta (\mathbf{x})$.

\begin{equation}
\mathcal{L}_{in} = || T^{-1}_\theta(\mathbf{x}) - T^{-1}_\theta(\mathbf{x}+\eta) ||_2
\end{equation}

\begin{table}[h]
	\centering
	\begin{tabular}{l l}
		\textbf{Parameter} & \textbf{Value} \\
		\hline
		\# levels & 3\\
		\# flow blocks per level ${N_f}$ & 4\\
		Coupling transform $C_{inter}$ & 256\\
		Base distribution $p(\mathbf{u}_1|\mathbf{h}_1)$, $p(\mathbf{u}_2|\mathbf{h}_2)$ & $\mathcal{N}(\mu(\mathbf{h}_i),\ Diag(\sigma(\mathbf{h}_i)))$\\
		Base distribution $p(\mathbf{u}_0)$ & $\mathcal{N}(\mu,\ Diag(\sigma))$\\
		Context Encoder $p(\mathbf{u}_1|\mathbf{h}_1)$, $p(\mathbf{u}_2|\mathbf{h}_2)$ & $N=5$ \\
		optimizer & Adam\\
		learning rate & $10^{-4}$ \\
		batch size & 15 \\
		\# steps & $20^5$ \\
		max gradient value & $10^5$\\
		max gradient $L_2$-norm & $10^4$\\
		latent noise magnitude &  $\pm 0.5$ \\
		latent noise loss ($\beta_{ln}$) & 100 \\
		autoencoder loss ($\beta_{ae}$)& 1 \\
		image noise loss ($\beta_{in}$) & 100\\
		image noise magnitude &  $\pm 10$ \\
		\hline
	\end{tabular}
	\caption{DIV2K training specification. }
	\label{table:div2k_training_params}
\end{table}

\begin{figure}[h]
	\centering
	\includegraphics[width=0.5\textwidth]{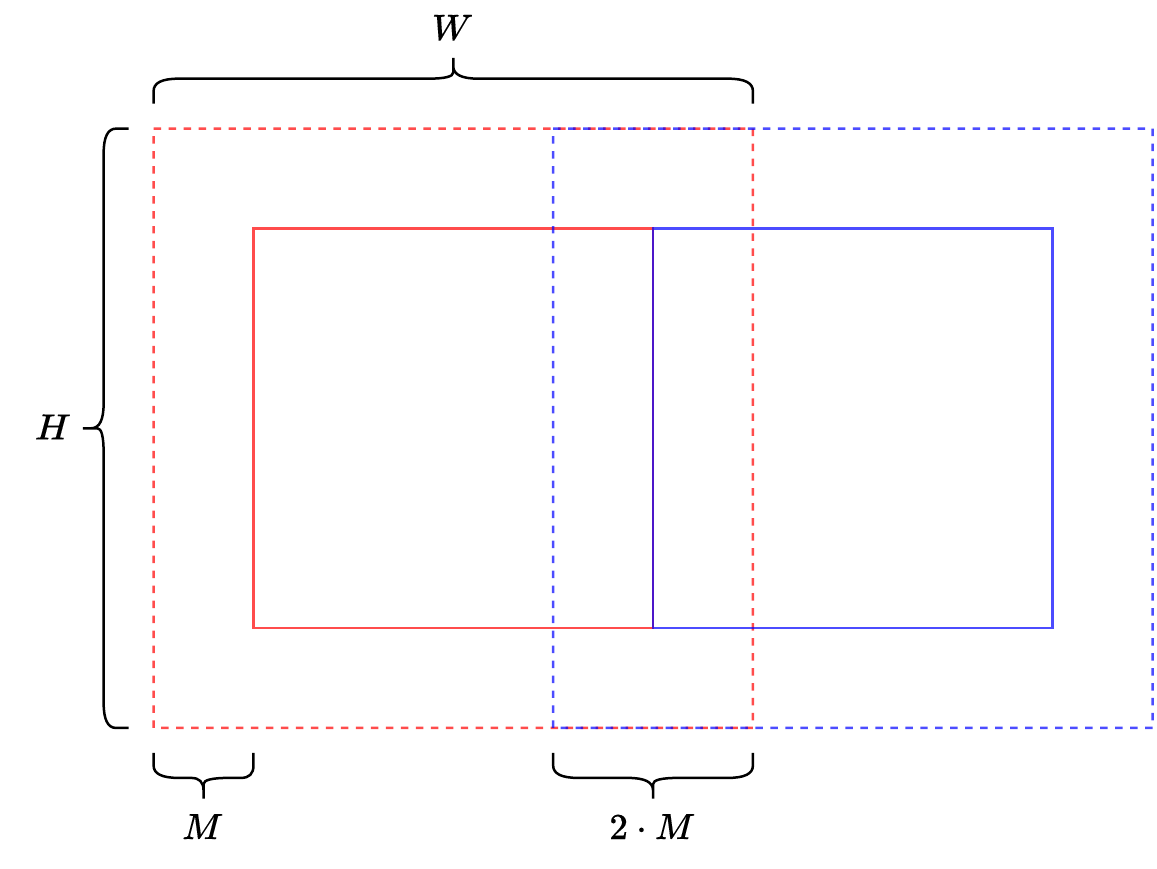}
	\caption[Tile margin illustration]{Illustrations of the tiles used for patch-wise reconstruction. $H$ and $W$ refer to the patches height and with respectively. $M$ refers to the margin. Neighboring patches overlap in a region of width $2\cdot M$. Analogously the same pattern extends in the vertical direction. In our work we use $H=W=64$ and $M=4$.}
	\label{fig:div2k_tile_margin}
\end{figure}

\paragraph{Patch-wise Reconstruction.}
A full image of arbitrary size can be reconstructed by reconstructing each patch individually. 
To avoid boundary artifacts between patches a margin is used as illustrated in Figure~\ref{fig:div2k_tile_margin}. 
The margin causes overlap between adjacent patches yielding more consistent results in boundary regions. 

\end{document}